\documentclass[epj]{svjour}
\usepackage{graphics}
\usepackage{color}
\newcommand{\ud}{\mathrm{d}}

\newcommand{\be}{\begin{equation}}
\newcommand{\ee}{\end{equation}}
\newcommand{\bea}{\begin{eqnarray}}
\newcommand{\eea}{\end{eqnarray}}

\begin{document}
\title{Collaborate, compete and share}
\author{Emanuele Pugliese\inst{1} \and Claudio Castellano\inst{2,1} \and Matteo Marsili\inst{3} \and Luciano Pietronero\inst{1,4}}
%
%
\institute{Dipartimento di Fisica, ``Sapienza'' Universit\`a di Roma,
Piazzale A. Moro 2, 00185 Roma, Italy
\and SMC, INFM-CNR, Piazzale A. Moro 2, 00185 Roma, Italy
\and The Abdus Salam International Centre for Theoretical Physics,
Strada Costiera 11, I-34014, Trieste, Italy
\and ISC, INFM-CNR, via dei Taurini 19, 00185 Roma, Italy}

\date{Received: date / Revised version: date}
%
\abstract{We introduce and study a model of an interacting population
of agents who collaborate in groups which compete for limited
resources. Groups are formed by random matching agents and their worth
is determined by the sum of the efforts deployed by agents in group
formation. Agents, on their side, have to share their effort between
contributing to their group's chances to outcompete other groups and
resource sharing among partners, when the group is successful. A
simple implementation of this strategic interaction gives rise to
static and evolutionary properties with a very rich phenomenology. A
robust emerging feature is the separation of the population between
agents who invest mainly in the success of their group and agents who
concentrate in getting the largest share of their group's profits.}

%

\maketitle
\section{Introduction}
\label{intro}

The collective behavior of a population of interacting individuals
often exhibits surprising properties, which can hardly be anticipated
on the basis of the micro-economic interaction. Game theory provides a
unified background for analyzing the outcomes of strategic
interactions among individuals rationally pursuing their
self-interest\cite{rubinstein94}.  Its
predictions, however, are often contradicted by empirical and
experimental research already in simple cases such as the prisoner's
dilemma or the ultimatum game \cite{prisoner,guth82,nowak00}.
This casts even more
serious doubts on its validity in cases where individuals face a more
complex strategic problem, involving a large number of other agents,
uncertainty and limited information.  On the other hand, many
collective phenomena in the natural sciences owe their peculiarity to
laws of statistical nature rather than to their specific microscopic
details. This suggests that an approach similar to that of statistical
physics can complement the game theoretic approach and highlight the
role of statistical laws in collective socio-economic phenomena. Much
work has been done along these lines on models of
socio-dynamics\cite{castellano08}
and on simple models of evolutionary dynamics \cite{nowak06,szabo07}.
For example, in spite of the doom predictions of game theory on the
inevitability of defective behavior in contexts such as the prisoners
dilemma, a different approach has shown that cooperative behavior and
altruism can indeed be sustained in a complex interactive environment,
in several ways~\cite{nowak06b}. 

Here we pursue this approach in a case where competition and the needs
to cooperate are intertwined in the strategic interaction. Our setup
is one where agents have to match in small groups, which compete for
limited resources. Agents in those groups which succeed in this
competition face the additional problem of sharing the profits among
themselves. Agents have to decide how to divide their effort in either
contributing to the success of their group or, when their group is
successful, in securing the largest share of the group's profit for
themselves. A concrete example of this is public funding of academic
research, for which empirical analyses start to appear~\cite{barber06}. 
Individual researchers form networks which submit proposals
for a specific call in their field. Only the best proposals get
accepted. Once a particular proposal is accepted the funds (and the
workload) are shared among the network partners. Each researcher may decide
to either commit a large effort to the preparation of projects,
which implies little effort in the negotiation stage if the project is
approved, or to commit a limited effort, thus ensuring a larger profit
when funds are to be divided but also making the whole project
weaker. This set up combines the incentives to free ride inside the
group with the necessity of cooperating with group members in order to
out-compete other groups. The former is akin to incentives to
defection in prisoner's dilemma or ultimatum games, whereas the latter
is typical of competitive behavior in markets, which is often
conducive to efficient outcomes (optimal resource allocation). Does
the tension between these two elements leads to virtuous or collusive
behavior? This is the key issue we address here.

We shall focus on a model which offers a simple realization of the
generic setup discussed above. This will not allow us to derive a
general answer to the question above, but rather illustrates the
complexity, i.e. the richness of behaviors, which lies behind it.
Still, by extending our basic model in several directions, we
shall be able to argue that our results are quite robust with
respect to simple modifications of the model.

The paper is organized as follows. In Sec.~\ref{model} we introduce
two versions of a simple model and discuss their behavior when
agents have fixed strategies.
The rest of the paper is focused on the second version of the model
(two-stage game). The constraint of fixed strategies
is lifted in Sec.~\ref{imitation}, where the complicated
phenomenology induced by imperfect imitation is described.
The possibility for agents to mutate (i.e. to adopt a totally new
strategy) is introduced and studied in Sec.~\ref{mutations}.
The effects of the variation of some features of the model
are described in Sec.~\ref{variation}.
Finally Sec.~\ref{conclusions} presents some conclusions.

\section{The model}
\label{model}

Let us first introduce an extremely simple model (possibly the simplest)
that describes the problem we are interested in.
Simple considerations about its phenomenology lead to the formulation
of a slightly more complicated model, that will be investigated in
the rest of the paper.

We consider the cooperation of $N$ agents to develop a scientific
project and the subsequent negotiation stage to share the awarded
grant\footnote{$N$ is chosen such that $N/N_p$ and $P/P_0$ are integers}.
Each project involves $N_p=2$ partners. At each time step every
agent participates to one and only one project.
There are hence $P=N/N_p=N/2$ projects.
Each agent is endowed at each time step with a total amount of effort
equal to 1 that he has to divide in two parts:
\begin{itemize}
\item A fraction $0 \leq E_i \leq 1$ is used for the development of the
project.
\item A fraction $1-E_i$ is used for the negotiation process to share
the grant awarded to the project.
\end{itemize}



At each time step agents are paired randomly and the resulting
projects are ranked depending on their quality, measured by $E_p$, the
sum of the energies devoted by participants to its development.
$E_p$ will then be smaller than $2$ (smaller than $N_p$ in the general case).
The $P_0$ projects with highest $E_p$ are financed with one unit
of payoff. 
Others are not financed, i.e. they do not get any payoff.
The threshold energy $E_s$ is the value of $E_p$ of the last project
financed. In the case that several projects have the same energy
equal to $E_s$ a random selection among them is performed.
Unless specified otherwise we will always take $P_0=P/2$.

Participants to successful projects negotiate then the division of the
grant awarded.
The assumption is that the share they receive is proportional to
effort they have not spent in the development of the project, that is
\be
\frac{1-E_i}{2-E_p},~~~~E_p=\sum_{j\in p} E_j,
\label{sharing}
\ee
where the sum over $j\in p$ includes all participants to the project $p$.
It is clear that there are (at least in principle) opposite drives
for each agent: is it
better to maximize the probability of success of the project
(taking $E_i$ large)
or the share obtained if the project is successful (taking $E_i$ small)?
As we shall see, much depends on the inter-temporal structure of the game.

\subsection{One stage game} 

It is relatively easy to understand that if each round is considered
independently then the best strategy of agents is to devote all their
efforts to proposing good projects if we consider the usual
assumptions of Game Theory about rationality of agents.

In order to see this, let us observe that the outcome of the game for
an agent with strategy $E$, and in particular his expected payoff
$\pi(E)$ depends on the distribution $f(E)$ of the energies of all
the others.  Assuming such distribution to be fixed, it is easy to
derive exactly, in the $N=\infty$ limit, the threshold effort $E_s$,
i.e. the minimal effort for a project to be financed, and using such a
value, the probability $P^\#(E)$ for a single agent to be financed. In
turn, this allows to compute the payoff $\pi(E)$.

Without entering into details, it is clear that if the threshold
effort $E_s$ is larger than 1, agents with effort $E < E_s -1 $ have
no chance to participate to a successful project, hence their success
probability and expected payoff will be strictly zero.  On the other
hand, if $E_s < 1$, ``zealous'' agents with $E>E_s$ will always have
their projects approved, independently on the effort of their
partner. Similar arguments suggest that, for a generic distribution
$f(E)$, the maximum expected payoff occurs for intermediate values
$\bar E$ of the effort: a careful balance between work and negotiation
is better than the extreme strategies $E=0$ and $E=1$.  Indeed the
former will too often fail to have their project approved whereas the
latter will derive very little benefits from the projects they take
part in.  If there is a best strategy $\bar E$, then all agents will
adopt it, suggesting that $f(E)$ should be sharply peaked around $\bar
E$. In this situation, it is clear that agents having a slightly
larger value of the effort $E=\bar E+\zeta$
(with $\zeta$ small enough) will have a larger
chance of getting the project through without paying too much in the
sharing phase, unless $\bar E=1$. Therefore, the situation with all
agents choosing strategy $E=1$ is the unique Nash Equilibrium of the
game. In other words, if agents can modify their strategies according
to a sensible evolutionary dynamics~\cite{nowak06} they will converge
to this virtuous asymptotic state.

This strategy is an Evolutionary Stable State:
if everyone has $E=1$, that is also the best strategy and no other
strategy can invade it.
It is also, in practice, an invading strategy itself, as
in Ref.~\cite{nowak90}, if we extend in a reasonable way the 
concept to our game:
any possible starting condition will tend toward this state.
In the one-stage setting, agents are then expected to converge to a 
simple symmetric equilibrium. Numerical simulations confirm this
expectation.

\subsection{Two stages game}

\begin{figure}
\resizebox{1\columnwidth}{!}{%
  \includegraphics{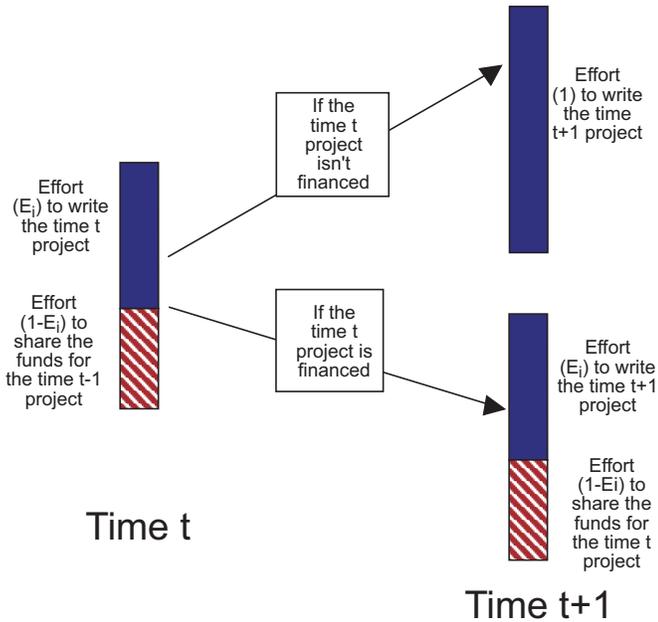}
}
\caption{Schematic representation of the division of the total effort of 
an agent in the two stages game.}
\label{figmodel}
\end{figure}

One unsatisfactory aspect of the discussion above is that agents who
are not financed stay idle and waste the effort $1-E$ they intended to
spend in the negotiation process. This might not be realistic in many
contexts. For this reason, we shall now move to a modified model based
on the idea that an agent that is not busy with an accepted project
will use all his effort in preparing the next one.  More in detail,
the duration of a project spans now two temporal units, as in
overlapping generation models \cite{OLG}: projects which are
``prepared'' in time step $t$ are ``run'' at time $t+1$. So at each
time there is a batch of projects which agents prepare for the next
period, and the previous batch of projects the best of which are
operational. From the agents' viewpoint, each has to share his unit of
effort between preparing new projects and sharing the profits, if they
are engaged in successful ones (Fig.~\ref{figmodel}).  The key
difference with the model above is that agents whose projects were not
financed in the previous step are not involved in any negotiation, so
they can use their whole effort endowment for the preparation of the
next project (this is somehow similar to the War of Attrition 
game with implicit time cost as in Ref.~\cite{eriksson04}, where the cost 
of waiting is just the inability to join other games).  Therefore the 
effort $E^{\rm eff}_i$ available to an agent for working at time 
$t+1$ is equal to $E_i$ or $1$ depending on whether at time $t$ his 
project was financed or not.

It is immediately clear that, at odds with what occurred in the
one-stage game, no value of $E$ is too low for an agent to enter
financed projects, because the unit effort available to an agent after
a failure gives him good chances of being successful at the next step.
This consideration intuitively leads to the expectation that there
will be basically two potentially favorable strategies: either work
always hard, have many financed projects and yield during the sharing
process (finite $E$) or work hard only every second project, have only
half of the projects financed but get as much as you can out of these
(vanishing $E$).

\begin{figure}
\resizebox{1\columnwidth}{!}{%
  \includegraphics{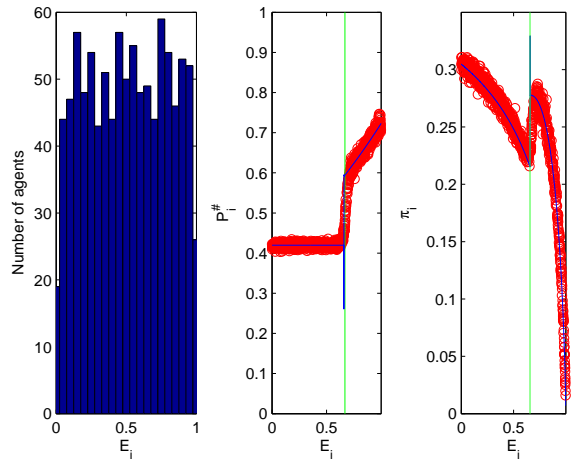}
}
\caption{\label{fig:2t} Behavior of the success probability $P^{\#}_i$ and
of the payoff $\pi_i$ for the agent $i$ (following strategy $E_i$) for a
fixed effort distribution $f(E)=1$;
the blue lines are the expected values (as computed in the Appendix),
the green vertical lines denote $E_s-1$.}
\end{figure}

Indeed, if the effort distribution function $f(E)$ is fixed, the
expected payoff $\pi(E)$ has two peaks, as shown in
Fig.~\ref{fig:2t} for a representative example.

For $E>E_s-1$ agents have a finite chance of being financed
even if they have been financed the time before.
For $E<E_s-1$ instead, agents can participate to a successful project
only when their effective effort is 1, i.e. they have not been financed
immediately before. As a consequence $P^{\#}(E) \le 1/2$.
In this class there is no advantage in having effort $E$ larger than zero,
because the probability of being financed is independent of $E$ and
$E>0$ is not beneficial in the grant division stage: this explains the
relative maximum of the payoff for $E=0$.
It turns out that very generally the same kind of phenomenon occurs also
for $E>E_s-1$ so that the payoff exhibits a second maximum for $E=E_s-1$.
Which of the two maxima is highest depends on the detailed form of $f(E)$.

A semi-analytical computation of the expected payoff for the two-stage
model is possible but non particularly insightful. Hence we shall skip it 
and move to the dynamics. In the appendix a numerical algorithm 
to compute the expected payoff and success probability is reported.

\section{Imitation dynamics}
\label{imitation}

With the goal of describing in a more realistic way a system of agents
competing and collaborating for obtaining and sharing grants, we now
add to the model an evolutionary dynamics, driven by imitation.  With
probability $r_E$ per unit time an agent has the possibility to copy
the strategy of another randomly chosen agent, if the latter has
higher cumulated payoff.  The imitation is imperfect, i.e. the new
effort of the imitating agent has a small random contribution between
$-\epsilon$ and $\epsilon$.

\begin{figure}
\resizebox{1\columnwidth}{!}{%
  \includegraphics{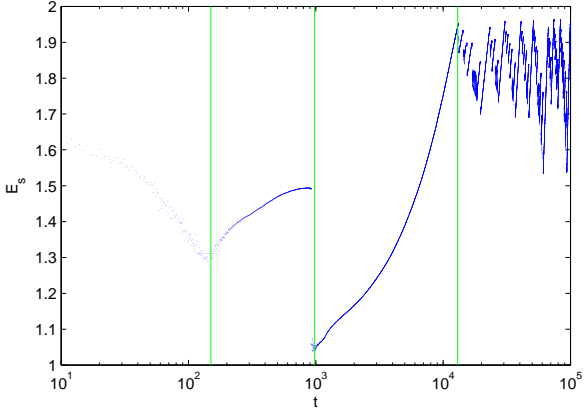}
}
\caption{\label{total_evolution} 
Temporal evolution of the threshold effort $E_s(t)$ for a system
of $N=10000$ agents and imitation rate $r_E=0.04$.
The initial form of the effort distribution is $f(E)=1$.
The vertical lines help to distinguish the four different regimes.}
\end{figure}

Numerical simulations show that in this case the behavior of the
system is quite rich.
The temporal behavior of the threshold effort $E_s$
(Fig.~\ref{total_evolution}) shows clear signatures of 
four well distinct temporal regimes, that reflect different
shapes of the effort distribution $f(E)$.

\begin{figure*}
\resizebox{2\columnwidth}{!}{%
  \includegraphics{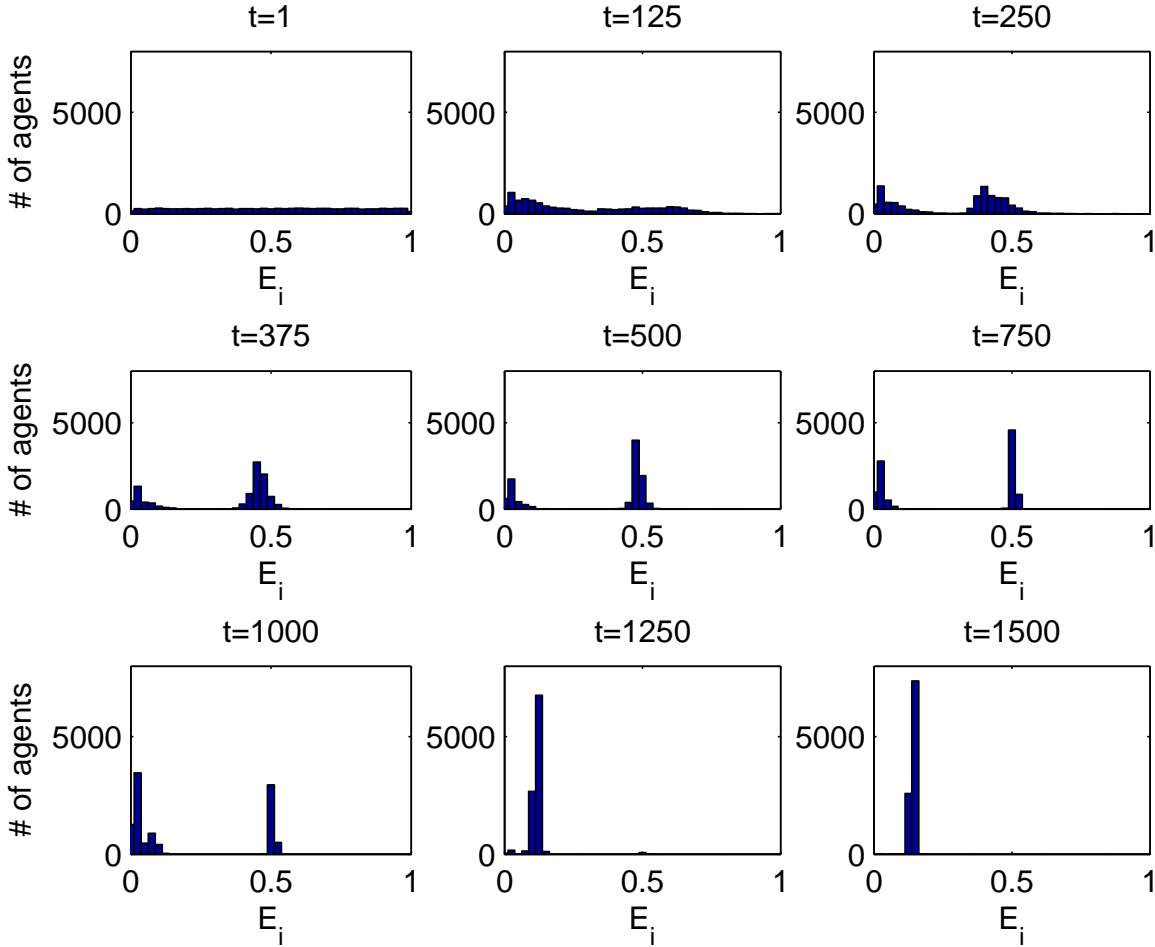}
}
\caption{\label{fig:2picchi} 
Histograms of the effort distribution $f(E)$ during the first two regimes. 
The initial $f(E)$ is uniform, $r_E=0.04$ and $N=10000$.
The formation of the two peaks is followed by their competition
leading to the disappearance of the peak in $E \approx \eta$.}
\end{figure*}

\begin{enumerate}
\item
Starting from any initial effort distribution $f(E)$, two narrow peaks
are formed, one for $E \approx 0$ and the other at a value $\eta$ of
the effort $E$.
\item
The two peaks evolve and compete, until the peak in $\eta$ disappears.
\item
The remaining peak drifts towards high values of $E$.
\item
It eventually enters
a sort of stationary state with intermittent oscillations about a large
value of $E$.
\end{enumerate}

This phenomenology occurs for any value of $r_E>0$, provided it is much
smaller than 1, otherwise the system becomes completely random.
The dependence on the initial shape of $f(E)$ is weak and does not
change qualitatively the picture.
We now describe in more detail the four regimes.

\subsection{Initial regime: formation of two peaks}

As intuitively expected from the static version of the model,
agents will tend to imitate other agents with energies corresponding
to the payoff maxima (see Fig.~\ref{fig:2t}). This quickly leads to
the collapse of the distribution function $f(E)$ in two peaks, one
around 0 and the other around a finite value $\eta$ not far from the
initial value of $E_s-1$ (see the first four panels of Fig.~\ref{fig:2picchi}).
At the same time, since the effort distribution varies,
$E_s$ slowly changes in time (Fig.~\ref{total_evolution}).

\subsection{First intermediate regime: competition of the two peaks}

When there are two peaks the dynamics can be monitored by two 
time-dependent quantities:
\begin{itemize}
\item
$\eta(t)$: the average position of the second peak (the
first is always around $E=0$).
\item
$\rho_0(t)$ the fraction of agents belonging to the first peak
(obviously $\rho_\eta=1-\rho_0$).
\end{itemize}

\begin{figure}
\resizebox{1\columnwidth}{!}{%
  \includegraphics{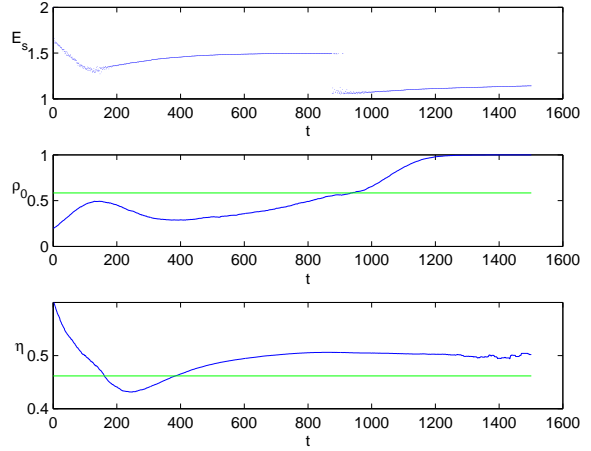}
}
\caption{\label{rho_mu_competition} 
Behavior of the threshold energy $E_s$, the amplitude $\rho_0$ of the peak
in $E=0$ and the position $\eta$ of the other peak, as a function of time
in the first and second regime for a system with parameters $r_E=0.04$
and $N=10000$. The horizontal lines are for the values
$\rho_0=7/12$ and $\eta=6/13$ (derived in the Appendix), that separate
different regimes.}
\end{figure}

Numerically, one observes that the peak in $\eta$ tends to move towards
the right (and upward in Fig.~\ref{total_evolution}), while 
simultaneously its total population decreases (i.e. $\rho_0$ grows).
This is evident from Fig.~\ref{fig:2picchi}, where the evolution
of the effort distribution $f(E)$ is represented, and from
Fig.~\ref{rho_mu_competition}, where $E_s$, $\rho_0$ and $\eta$ are reported
as a function of time.

The drift of the second peak towards larger energies is easily understood
qualitatively. The fundamental observation is that the threshold effort
$E_s(t)$ falls within the peak in $\eta$, which has a width of the order
of $\epsilon$. As a consequence, agents with energies in the left part of the
peak will be less financed than those in the right part and then will
tend to imitate them. The peak position slowly drifts towards right because
of the struggle of agents to work just a little bit more than their
colleagues.

The dynamics of $\rho_0(t)$ can be understood by
making use of the simplifying assumption that the peaks are
Dirac-$\delta$ functions.
Under such assumption one can compute the expected payoff of agents in
the two peaks as a function of $\rho_0$ and $\eta$ (Fig.~\ref{payoff}).
See the appendix for details about the derivation.

\begin{figure}
\resizebox{1\columnwidth}{!}{%
  \includegraphics{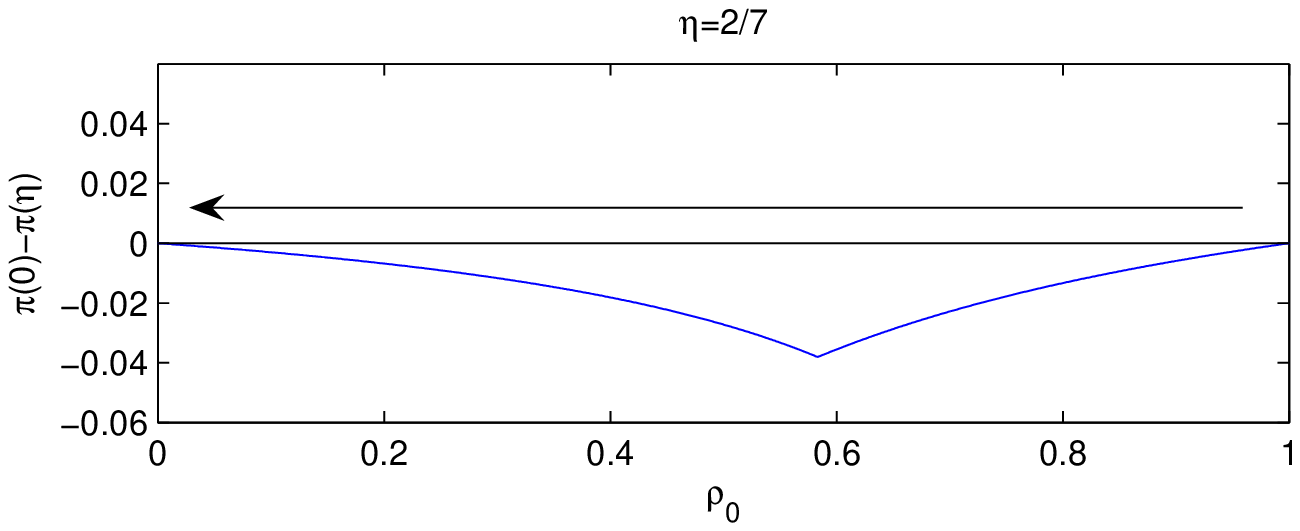}
}
\resizebox{1\columnwidth}{!}{%
  \includegraphics{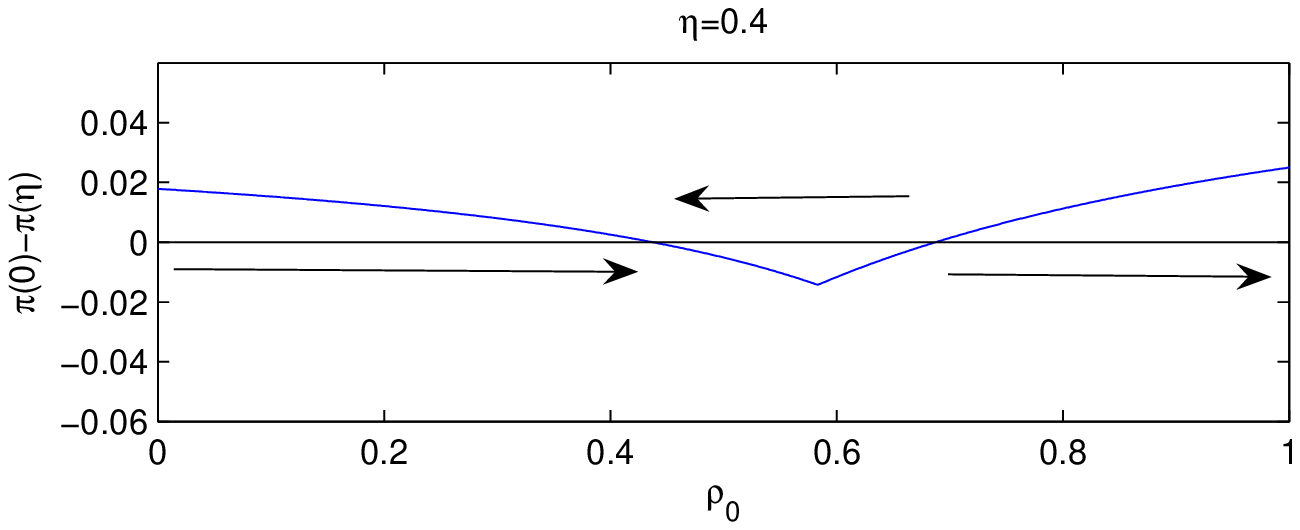}
}
\resizebox{1\columnwidth}{!}{%
  \includegraphics{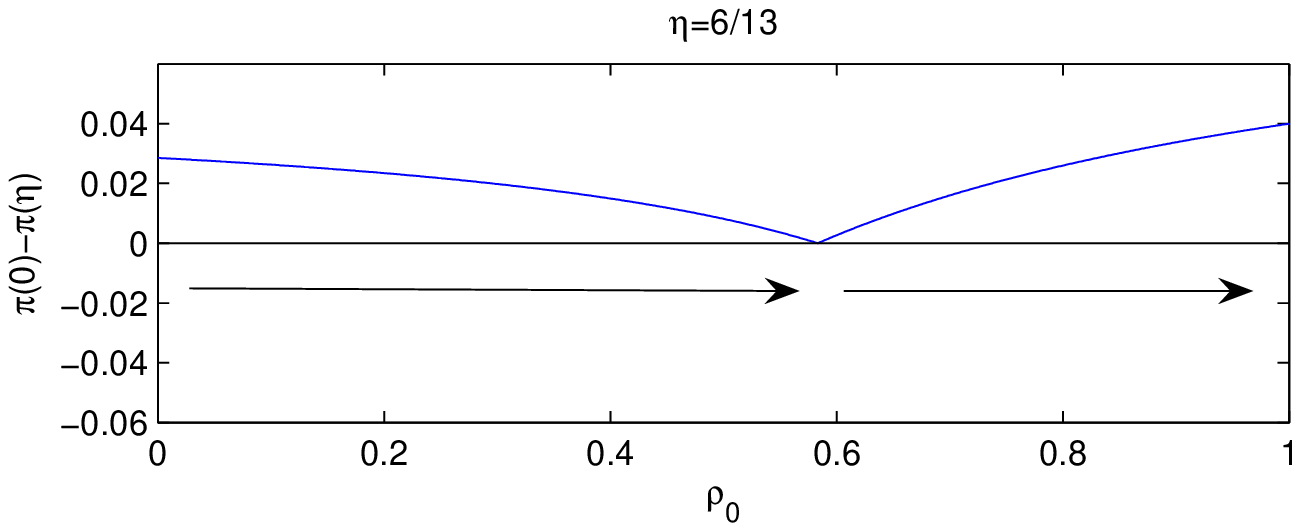}
}
\resizebox{1\columnwidth}{!}{%
  \includegraphics{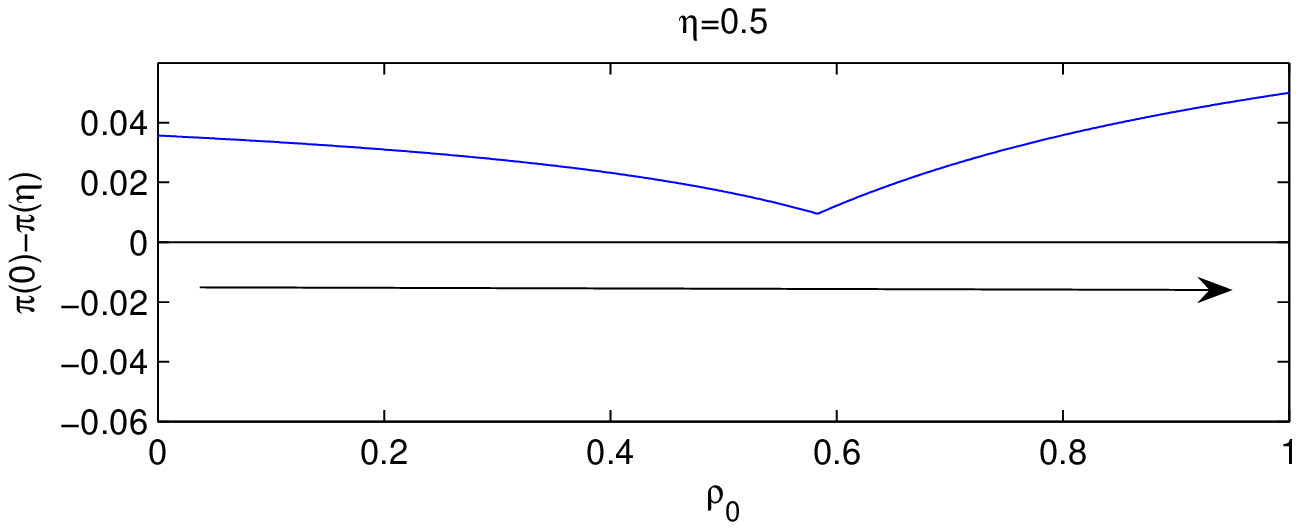}
}
\caption{\label{payoff} 
Behavior of the payoff difference $\pi(0)-\pi(\eta)$ between an agent with
effort $E=0$ and one with effort $E=\eta$ as a function of $\rho_0$
for different values of $\eta$. The arrows indicate the direction of the
dynamical evolution.}
\end{figure}

For small $\eta<2/7 \approx 0.286\ldots$ the strategy $E=\eta$
is more convenient than $E=0$ for any value of $\rho_0$: agents
in $\eta$ will be imitated and $\rho_0$ decreases.

For $2/7<\eta<6/13$ there are two values of $\rho_0$ for which the
payoff of the two peaks is the same.
\footnote{The asymptotic value of $\eta$ is actually larger than $6/13$,
probably due to the effect of the finite width of the two peaks.}
The dynamics of $\rho_0(t)$ is much faster than the evolution
of $\eta(t)$, as a detailed evaluation brings out
that the rate of change ${\dot \eta}$ is
smaller than ${\dot \rho}$ by a factor of the order of $\epsilon$.
As a consequence, for any $\eta$, $\rho_0(t)$ assumes the
stable ``equilibrium'' value corresponding to the leftmost point where the two
payoffs are the same. The temporal evolution of $\rho_0(\eta)$ is then
enslaved by the evolution of $\eta(t)$.

Finally, for $\eta>6/13$, the payoff of agents in zero is always larger
than the payoff of agents in $\eta$. Therefore when this critical
value is reached (corresponding to $\rho_0=7/12$) an abrupt change
occurs: the threshold effort suddenly moves towards 1,
the relative balance between the two peaks breaks
down and rapidly the peak in $\eta$ disappears.

Notice that the passage from $E_s$ of the
order $1+\eta$ to a value close to 1 is a rather intermittent process
(Fig.~\ref{rho_mu_competition}):
due to fluctuations associated with the random pairing of agents and
the finiteness of $N$, for some time the threshold energy bounces back
and forth between the two values before setting to a value close to 1.

\subsection{Second intermediate regime: single peak drifting}
Once a single peak is present, its position drifts towards larger effort
values (Fig.~\ref{total_evolution}) for reasons perfectly analoguous
to what happened before to the peak
in $\eta$: agents in the right part of the peak tend to be more successful
and hence be imitated.
In this ``ideal'' regime agents constantly improve themselves by
devoting more and more effort to projects and less and less to the
negotiation stage.

\subsection{Asymptotic regime: strong intermittent oscillations}

Also the ideal stage is doomed to end.
This happens when the width of the peak $\Delta E^*$ becomes comparable to
$1-E^*$, the distance from 1 of the peak position $E^*$.
In such a case, a strategy in the extreme left tail of the peak becomes
convenient, because the reduced chance of being financed is compensated by
a strong advantage in the grant division. This leads to the formation
of a secondary peak for $E \approx  E^*- \Delta E^*$.
At this point a competition between two peaks similar to the one of
the First intermediate stage takes place, leading to a succession of cycles
where the rightmost peak disappears, the remaining drifts towards 1 and 
then splits again in two peaks.

\section{The role of mutations}
\label{mutations}

We now consider the effect induced by the
possibility of agents to mutate, i.e. to adopt a new strategy between
$E=0$ and $E=1$ in a completely random way.
The fraction of mutating agents per time step $r_M$ is taken to be much
smaller than the imitation rate $r_E$, so that mutations do not completely
change the general dynamics except for the possibility of agents to explore
the whole space of strategies. In other words, they make the dynamics
innovative, still preserving monotonicity with respect to the
payoff~\cite{rubinstein94}.

While the general picture remains the same, with the competition between
two types of strategies, mutations modify the
detailed phenomenology. In particular they make states with
a single peak always unstable, thus preventing stages 3 and 4 of the
mutationless case (second intermediate and asymptotic regimes)
to be reached.
After the initial formation of two peaks, the competition between them
leads, as before, to growth of the peak in 0 and to the 
fast decay of the one in $\eta$.
At this point it is necessary to distinguish between two possible cases.

\begin{figure*}
\resizebox{2\columnwidth}{!}{%
  \includegraphics{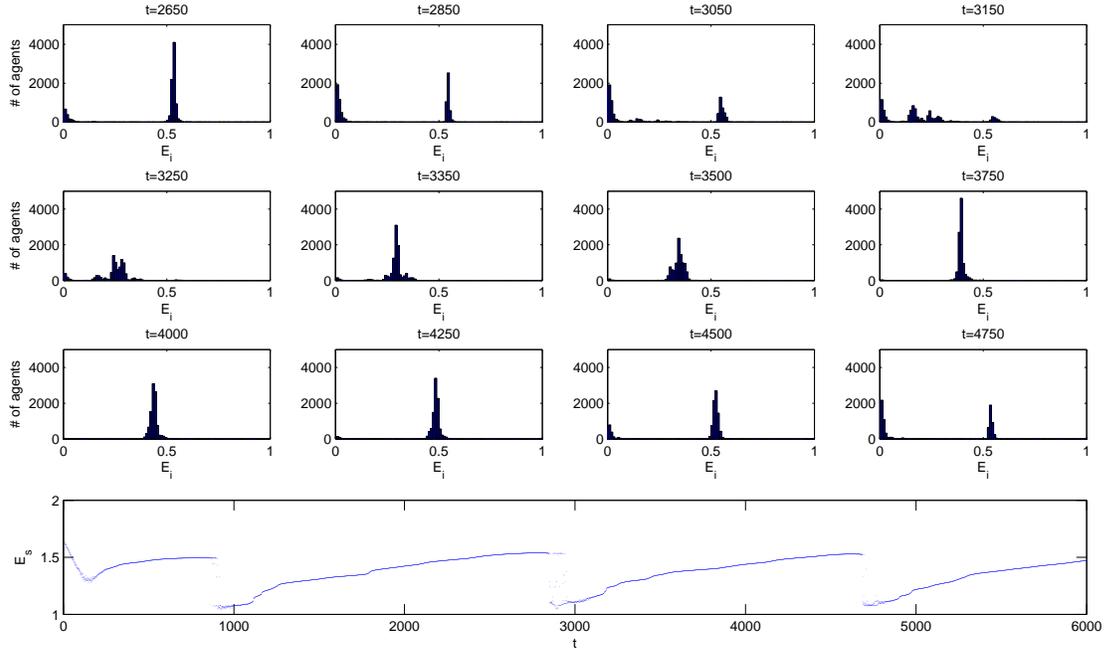}
}
\caption{\label{fig:longMut}
Temporal evolution of the histogram of the effort distribution $f(E)$ and of
the threshold energy $E_s$ for a system with $r_E=0.04$, $N=10000$ 
and a very small mutation rate $r_M = 0.0001$.}
\end{figure*}

If $r_M$ is extremely small with respect to $r_E$ (see below for 
details) mutated agents probe all possible effort values between
$0$ and $\eta$ and one of them gives rise to a new peak that rapidly
overcomes the others and starts drifting right. Once it goes beyond
$\eta=2/7$, as it can be seen from Fig.~\ref{payoff}, the strategy $E=0$
becomes again advantageous against agents in $\eta$ (with $\rho_0=0$). 
A peak in zero starts to grow, while $\eta$ grows further, 
bringing the system back to the first intermediate regime.
The system enters therefore in a cycle with two peaks that 
cyclically collapse and reappear.
(Fig.~\ref{fig:longMut}).

\begin{figure*}
\resizebox{2\columnwidth}{!}{%
  \includegraphics{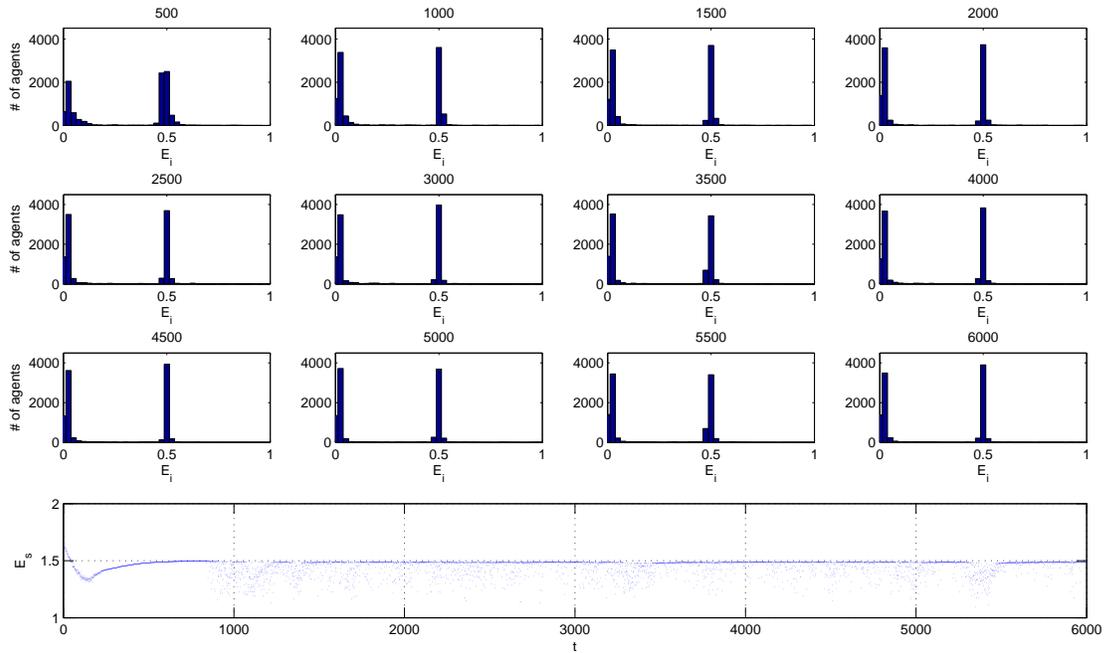}
}
\caption{\label{fig:longMut2}
Temporal evolution of the histogram of the effort distribution $f(E)$ and of
the threshold energy $E_s$ for a system with $r_E=0.04$, $N=10000$ 
and a small mutation rate $r_M = 0.001$.}
\end{figure*}

For larger values  of $r_M$ (but still $r_M \ll r_E$) instead,
the state with one peak in 0 and one in 0.5 becomes essentially
evolutionary stable (Fig.~\ref{fig:longMut2}).
The populations of the two peaks will be around the critical
values $\rho_0=7/12$, $\rho_\eta=5/12$. Depending on small variations
of such populations, and on the creation of some agents with energies
between them, $E_s$ will then fluctuate intermittently between $1$ to $1.5$

By considering different system sizes it turns out that the boundary
between these two behaviors goes to a finite limit
(of the order of $r_M/r_E \approx 10^{-2}$) when $N$ diverges.

\section{Robustness of the phenomenology}
\label{variation}

What happens if some of the assumptions underlying the model are changed?
In order to investigate the robustness of the phenomenology
presented above, we consider several modifications applied to the
model with both imperfect imitation and enough mutations to generate
a stationary state with two peaks in equilibrium.

If the fraction of profits received by an agent, Eq.~(\ref{sharing})
is taken proportional to $(1-E)^{\alpha}$, with $\alpha>0$, simulations
show that no qualitative change occurs with respect to the case
$\alpha=1$, studied above. 

Another possibility is to introduce some stochasticity in the probability
to be approved. So far the probability for a project of total effort
$E_p$ to be financed has a sharp threshold:
$P(E_p>E_s)=1$, $P(E_p<E_s)=0$.
If we consider instead a smooth function
\be
P(E_p) = \frac{1}{\exp \left[ (E_p-E_s)/T\right] + 1},
\ee
it becomes possible that a project of relatively small value is financed
while a better one (higher $E_p$) is not. Simulations
again show that nothing changes qualitatively in the global behavior of the
system, provided the ``temperature'' $T$ is not too high.

\begin{figure}
\resizebox{1\columnwidth}{!}{%
  \includegraphics{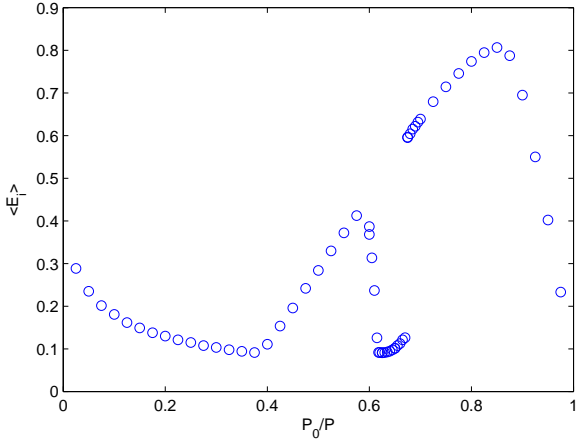}
}
\caption{\label{Q} Average effort of agents in the stationary state
as a function of $Q=P_0/P$.
System parameters are $N=1000$, $r_E=0.04$, $r_M=0.005$.}
\end{figure}

One parameter that turns out instead to affect noticeably the
results is the fraction $Q$ of approved projects.
So far we have always taken $P_0=P/2$, i.e. half of the
projects were financed and half of them were not.
For generic value of $Q=P_0/P$
projects can be distinguished in three categories.
If both agents taking part to the project were not financed at the
previous step then $E_p=2$.
This occurs with probability $(1-Q)^2$. If one of them was financed
(probability $2Q(1-Q)$) then $1 \le E_p \le 2$; else, with 
probability $Q^2$, $0 \le E_p \le 1$.
After a transient stage, whose duration may vary with $Q$,
the system settles in a stationary state.
The properties of such state depend on the value of $Q$ leading
to the identification of several different regimes,
reflected in the behavior displayed in Fig.~\ref{Q}.

For $Q\leq\frac{3-\sqrt{5}}{2}\approx 0.38$, since $(1-Q)^2>Q$
all financed projects
have effort $E_p=2$, i.e. they involve partners that have not been
financed at the previous step.
The probability of being financed does not depend on $E$.
Hence all agents converge to the $E=0$ strategy: too few financed
projects push agents to laziness and greedy negotiations.

For $\frac{3-\sqrt{5}}{2}<Q<\frac{\sqrt{5}-1}{2} \approx 0.62$ 
the behavior is of the same type of the case studied so far $Q=1/2$,
with the competition of two peaks, one in 0 and the other in $E=\eta$,
whose amplitude goes to zero at the boundaries of the range.

For $Q>\frac{\sqrt{5}-1}{2}$ the situation is more involved.
In general a single peak is formed rather than two.
For $Q \approx 0.7$ the peak jumps from an effort close to zero
to one larger than $1/2$, because the presence of agents with $E>1/2$
reinforces the profitability of such strategies.

It is important to remark that the presence of three distinct regions
is related to the number of participants to each project $N_p=2$.
For a generic $N_p$ there are $N_p+1$ zones, separated by points where
the $E=0$ strategy dominates all others. Variations on $N_p$ 
have also other effects, mainly quantitative: the more partecipants 
per project, the less the agents tend to work.
The second peak  will move toward $0$.
Of course when there are too many agents per 
project the two peaks eventually become indiscernible.

\section{Conclusions}
\label{conclusions}

In this paper we have introduced a very simple model for
the formation of collaborations in scientific projects and the ensuing
stage where partners negotiate to share the awarded grants.
The model includes only the main features of the process.
It can therefore be viewed as describing a generic
situation where agents have to collaborate within a group to compete
globally for limited resources but also negotiate within the
group to share the profit earned together.

In order to identify the most relevant effects and understand
their role in the resulting phenomenology we have kept the
description at a very basic level and perfomed an exploratory
investigation, with no claim of considering any realistic situation.
Despite these clear oversimplifications, the model
displays some interesting, if not realistic, patterns of behavior.

Generically, within the two-stage game, the population of agents
spontaneously clusters around two different strategies.  ``Hard
workers'' are characterized by relatively large values of the effort
they put in the preparation of proposals. They tend to maximize the
quality of their projects, so that a large fraction of them is
financed. Consequently they put less effort in the negotiation stage.
``Greedy'' agents cluster instead around the strategy $E=0$, implying
that they try to maximize their share when a grant is obtained.  To do
this they accept that their proposals are financed less frequently,
and work hard only if this does not decrease their ability to argue
about how to share the profits.  Depending on the details of the
model, the two populations are either in equilibrium or exhibit
oscillations, but the existence of two classes of behavior is a robust
feature.

Finally, the study of the dependence on the fraction of approved projects,
shows that a small number of rejected projects is sufficient to push
agents towards a state with minimum internal competition
and large productive effort.
On the other hand, when $Q$ is very small,
so that the time lag between approved projects is large, agents tend to
struggle for the scarce resources when they are available, instead of
committing themselves to productive effort.

These interesting findings clearly call for additional investigations.
In the spirit of going towards more realism, many possible variations
of the model may be devised.
One of the most natural is to consider
heterogeneity either in agents' intrinsic skills (the total effort
available) or in the composition of projects (number of participants).

A further interesting direction has to do with extensions to
interaction structures which are more constrained and realistic than
the random matching assumed here. Preliminary results show, for
agents placed on a square lattice, interesting
spatio-temporal patterns with phase segregation between a population
with $E\approx 0$ and $E\approx 1$, if both the ranges of imitation
and collaboration are limited.
A natural generalization of this has to
do with allowing agents to select their neighbors depending on past
performance. Preliminary results suggest that both constraining the
interaction pattern and allowing agents to selectively choose
neighbors generally increases the global performance of the system, as
measured by the threshold effort $E_s$ for a project to be
financed. The simultaneous evolution of the network of connections and
of the distribution of agent strategies is another promising direction
to follow. The systematic study of these effects deserves a systematic
study to be discussed elsewhere.

\section*{Appendix}

In this appendix we report the derivation of the payoff curves
shown in Fig.~\ref{payoff}.
This is carried out assuming that each peak has vanishing width, so that
it is mathematically described by a Dirac's $\delta$ function.
We take the first peak to be in $E=0$ and the second in $E=\eta$.
In this case the system is fully described by the quantities
$\rho_{0^-}$, the density of agents with zero effective effort, and 
$\eta$ that sets the effort of the other peak.
The superscripts $+$ e $-$ represent agents that commit in the project
an effective effort equal to $1$ or $E$, respectively.

At each time step the number of agents that does not receive a grant
is half of the total.
Hence $\rho_+=1/2$ and from the normalization condition 
$\rho_{0^-}+\rho_{\eta^-}+\rho_{+}=1$
one gets $\rho_{\eta^-}=\frac{1}{2}-\rho_{0^-}$.

The effort $E_p$ of a project, given by the sum of effective energies
of two randomly selected agents, can be
\begin{itemize}
\item $0$, with probability $\rho_{0^-}^2$
\item $\eta$, with probability $2 \rho_{0^-} (\frac{1}{2}-\rho_{0^-})$
\item $2\eta$, with probability $(\frac{1}{2}-\rho_{0^-})^2$
\item $1$, with probability $2 \frac{1}{2} \rho_{0^-}$
\item $1+\eta$, with probability $2 \frac{1}{2} (\frac{1}{2}-\rho_{0^-})$
\item $2$, with probability $\frac{1}{4}$
\end{itemize}

Projects with total effort $E_p=2$ are always financed. 
Which of the other projects gets funded (and hence what is
the threshold effort $E_s$) depends on the value of $\rho_0^-$ and $\eta$.

For $\rho_{0^-} \leq 1/4$ the projects with effort $E_p=1+\eta$ and $2$
are together more than half of the total number $P$.
As a consequence $E_s$ will be $1+\eta$.
Otherwise, if $\rho_{0^-}>1/4$, other projects are funded.
For $\eta<1/2$, that is the relevant case, the remaining projects
with higher effort are those with $E_p=1$, i.e. the projects with
one partner in state + and one in state $0^-$.

The probability to be financed, as a function of the effective
effort $E^{\rm eff}$ is then, for $\rho_{0^-} \leq 1/4$

\begin{eqnarray}
P^{\#}(E^{\rm eff}=0) &= &0 \\
P^{\#}(E^{\rm eff}=\eta) &=& \frac{1}{4-8 \rho_{0^-}} \\
P^{\#}(E^{\rm eff}=1) &=&\frac {3}{4}
\end{eqnarray}

In order to compute the expected payoffs
one needs to evaluate, for agents with strategy $0$ ($\eta$)
how many of them are in effective state $0^+$ ($\eta^+$) and how many
in state $0^-$ ($\eta^-$).
This is carried out by assuming that the populations in state
$+$ or $-$ are in equilibrium, i.e. the rate of transition from
$-$ to $+$ [$R(- \to +)$] is equal to the opposite rate $R(+ \to -)$.
This is reasonable because these processes are much quicker than all
others, and in particular imitation.
For the strategy $E=0$ this yields
\bea
R(- \to +) &= &\rho_{0^-} (1-P^{\#}(0^-))=\rho_{0^-} \\
R(+ \to -) &= &\rho_{0^+} P^{\#}(+)=\frac{3}{4}\rho_{0^+},
\eea
so that 
\be
\rho_{0^-}=\frac{3}{7}\rho_0 ~~~~~~~~~~~~~~\rho_{0^+}=\frac{4}{7}\rho_0.
\ee

Similarly, for $E=\mu$
\bea
R(- \to +) &= &\rho_{\eta^-}\frac{3-8\rho_{0^-}}{4-8\rho_{0^-}} \\
R(+ \to -) &= &\frac{3}{4}\rho_{\eta^+},
\eea
so that 
\be
\rho_{\eta^-}=\frac{3-6 \rho_{0^-}}{6-14 \rho_{0^-}}\rho_{\eta}
 ~~~~~~~~~~~~~~
\rho_{\eta^+}=\frac{3-8\rho_{0^-}}{6-14 \rho_{0 ^-}}\rho_{\eta}.
\ee
From the condition on $\rho_{0^-}$ one obtains that 
$\rho_{0^-}\leq\frac{1}{4}$ if $\rho_0\leq\frac{7}{12}$.

We are now in the position to determine the average expected payoffs $\pi(0)$
and $\pi(\eta)$, by summing, over all possible pairings, the probability
of such pairing times the associated expected payoff.

\bea
\pi(0)&=&\frac{21-8\eta\rho_0}{49(2-\eta)}\\
\pi(\eta)&=&\frac{98-84\rho_0+\eta(-49+32\rho_0^2)}{196(2-\eta)(1-\rho_0)}
\eea

The application of the same procedure with the appropriate values
for $P^\#(E^{\rm eff})$ gives the expression of the average payoffs also
for $\rho_{0^-} > 1/4$ ($\rho_0 > 7/12$). The behavior of the payoff
difference $\pi(\eta)-\pi(0)$
in the whole range of $\rho_0$ is plotted in Fig.~\ref{payoff}.

\paragraph{Computation of $P^{\#}(E)$ for generic fixed $f(E)$ for the two
stages game}

When $f(E)$ is not a sum of $\delta$ functions, the computation is done in an
iterative way.
Considering $f(E)$ fixed (or slowly evolving, this can be done in a wide
range of the parameters), we call $f_t(E)$ the varying effective distribution,
changing in time because agents go to a $+$ or $-$ status.
Starting from values at $t-1$, we compute values at $t$ in this way:
\begin{enumerate}
\item $f_{t}(E)=f(E) P^{\#}_{t-1}(E)+\frac{P_0}{P}\delta^-(E-1)$;
\item we convolve $f_{t}$ to obtain the probability density of the energy
of the projects at time $t$, $f^P_{t}(E)=\int_0^1f_{t}(E')f_{t}(E-E')\ud E'$;
\item we obtain $E_s^t$ as the value such that
$\int_0^{E_s^t}f^P_{t}\ud E'=\frac{P_0}{P}$;
\item at last, we compute
$P^{\#}_t(E)=P^{\#}_{t-1}(E)\int^1_{Es-E}f_t(E')\ud E')+(1-P^{\#}_{t-1}(E))\int^1_{Es-1}f_t(E')\ud E'$.
\end{enumerate}

This procedure allows to compute numerically $P^{\#}(E)$.
A stable function, in very good agreement with simulations, is obtained
after around 10 iterations. In Fig.~\ref{fig:2t} we used 80 iterations.

Once $P^{\#}(E)$ is found, it is easy to compute $\pi(E)$ considering
all possible combinations of $+$ and $-$.

\end{document}